\documentstyle[psfig]{mn}

\title[The dark halo of the lens galaxy]{The dark halo of the main lens galaxy 
in QSO 0957+561}
\author[Goicoechea et al.]
      {L.\ J. Goicoechea$^1$, R. Gil-Merino$^{1,2}$ and A. Ull\'an$^1$\\
	$^1$ Departamento de F\'{\i}sica Moderna, Universidad de Cantabria,
Avda. Los Castros s/n, E-39005 Santander, Spain \\ E-mail: 
goicol@unican.es, aurora.ullan@postgrado.unican.es\\
	$^2$ Institute of Astronomy, School of Physics, University of Sydney,
NSW 2006, Australia \\ E-mail: rodrigo@physics.usyd.edu.au}	

\begin{document}
\onecolumn

\maketitle

\begin{abstract}
We present an analysis of infrared/optical/ultraviolet spectra of the two images 
of the first gravitationally lensed quasar Q0957+561A, B. The Hubble Space 
Telescope observations of Q0957+561A and Q0957+561B are separated in time by the 
known time delay in this system, so we can directly deduce the flux ratios. These 
flux ratios of images lead to important information on the dark halo of the main 
lens galaxy (a giant elliptical at redshift $z$ = 0.36). Our measurements for the 
continuum are in good agreement with extinction in the elliptical galaxy and a 
small fraction of mass in collapsed objects (no need for gravitational 
microlensing). From the continuum and emission line ratios, we also show evidence 
in favour of the existence of a network of compact dusty clouds.    
\\
\\
This is a preprint of an Article accepted for publication in {\it MNRAS Letters}
$\copyright$ 2005 Royal Astronomical Society
\end{abstract}


\section{Introduction}

The populations of galaxy dark haloes may include collapsed objects (black holes, 
brown dwarfs, etc.) as well as non-collapsed structures and elementary particles
(e.g., Jetzer 1999). While some experiments suggest that the Milky Way dark halo 
is not dominated by collapsed objects with stellar or substellar mass (Alcock et 
al. 1998, 2000; Lasserre et al. 2000; Alcock et al. 2001), the populations of 
relatively far galaxies are still largely unknown. Therefore, extragalactic 
studies are crucial to reveal the nature of dark matter.

The double (gravitationally lensed) quasar Q0957+561A, B (at redshift $z$ = 1.41) 
was discovered 25 years ago in a radio survey (Walsh, Carswell \& Weymann 1979). 
At optical wavelengths, the main lens galaxy in the system (at redshift $z$ = 
0.36) appears as an extended source close to the image B. This giant elliptical 
galaxy is part of a cluster of galaxies that also contributes to the lensing 
(Stockton 1980; Garrett, Walsh \& Carswell 1992). The angular separation between 
the B image and the centre of the galaxy is only of $\approx$ 1 arcsec, whereas 
the angular separation between the A image and the galaxy is about five times 
larger (Bernstein et al. 1997). This translates in that the light rays associated 
with the images A and B are characterized by two different impact parameters,  
$\approx$ 18 $h^{-1}$ kpc and 3--4 $h^{-1}$ kpc, respectively, so the two beams 
are embedded in the galaxy halo and might unveil the structure of the dark matter
in the cD galaxy. The beams may be affected by gravitational microlensing by stars 
and/or collapsed dark objects (Chang \& Refsdal 1979), and/or extinction by clouds 
of gas and dust. Hence, the spectra and/or light curves of the quasar images could 
reveal the dark content of the galaxy. We consider the concordance model of the 
Universe (70\% of the cosmos is thought to be dark energy, $\Omega_{\Lambda}$ = 
0.7, and 30\% matter, $\Omega_M$ = 0.3), and $h$ is the normalized Hubble constant 
(0.5 $< h <$ 1).

When multiwavelength observations of the two images are separated in time by the 
known time delay in the system (e.g., Serra--Ricart et al. 1999), one obtains the 
flux ratios in a proper way (Schild \& Smith 1991). The standard gravitational 
scenario predicts the existence of an achromatic and stationary flux ratio of 
images, i.e., the macrolens flux ratio (Schneider, Ehlers \& Falco 1992), but the 
observations of multiple quasars often disagree with this standard prediction 
(e.g., Nadeau et al. 1991; Jaunsen \& Hjorth 1997; Burud et al. 2000). Flux ratio 
anomalies are thus basic tools to investigate lens galaxy haloes. In this paper 
(Section 2) we describe the relevant Hubble Space Telescope (HST) spectra for the 
two components Q0957+561A, B. In Section 3, we measure and interprete a large 
collection of infrared/optical/ultraviolet flux ratios. Finally, Section 4 
summarizes our main conclusions.

\section{HST spectra}

Spectra of Q0957+561A and Q0957+561B were obtained in 1999 April 15 and 2000 June 
2--3, respectively, with the Space Telescope Imaging Spectrograph (STIS) on board 
the HST. The spectra of each image cover a wide range of wavelengths, from the 
near--infrared (NIR) to the ultraviolet (UV), since the G230L, G430L and G750L 
gratings were used in the experiment. The central wavelengths of these gratings
are 0.24 $\mu$m (G230L), 0.43 $\mu$m  (G430L) and 0.77 $\mu$m (G750L). A 52 
$\times$ 0.2 arcsec$^2$ slit was also used in each observation. 

The final data (flux vs. wavelength) are calibrated with the last version of the 
CALSTIS pipeline software, so the corrections for time--dependent sensitivity of 
the MAMA detectors (which is relevant for the G230L observations) as well as for 
charge--transfer efficiency of the CCD detector (relevant for the G430L and G750L 
observations) are incorporated. We also note that the HST--STIS data of the two 
images do not incorporate any correction for cross--contamination of the spectra 
and contamination by the lens galaxy. However, the cross--contamination of the
spectra and the contamination by the galaxy light at the bluest wavelengths are
expected to be negligible. The possible contamination of Q0957+561B (due to its 
proximity to the galaxy) only must be checked at the reddest wavelengths. In the
next section, using complementary photometric observations, we test the HST--STIS
continuum flux ratios at 0.54--0.65 $\mu$m. 

\section{Flux ratios}

\subsection{Continuum emission}

We firstly focus on the continuum in the wavelength range 0.22-1 $\mu$m. The data cover 
the interval 0.09 $< \lambda_{qso} <$ 0.41 $\mu$m in the rest frame of the emitted 
radiation, i.e., blue/UV emission. This emitted radiation comes to the lens galaxy at 
0.16 $< \lambda_{gal} <$ 0.74 $\mu$m (optical/UV). We average the continuum flux over 
independent intervals of 100 \AA\ (0.01 $\mu$m) avoiding the prominent emission/absorption 
lines and bad pixels, and then infer 32 flux ratios $B/A$. The ratios appear in Figure 1 
(blue, green and red circles). For comparison to previous work, two additional black open 
circles are depicted in Fig. 1. These two optical continuum flux ratios (black open 
circles) present no contamination by the lens galaxy light, since the contamination in 
image B was accurately subtracted (Goicoechea et al. 2005; see also Ovaldsen et al. 2003;
Ull\'an et al. 2003). The corresponding frames in the $V$ and $R$ bands were obtained 
with the Nordic Optical Telescope in 2000 February/March and 2001 April. In  Fig. 1 there 
is an apparent agreement between the old results and the new measurements, so the previous 
data validate all the HST--STIS flux ratios at $x \ge$ 1.5 ($x$ = 1/$\lambda$, $\lambda$ 
in $\mu$m). However, as the contamination could be important at redder wavelengths (e.g., 
see the frames in the $H$ and $I$ bands at 
http://cfa-www.harvard.edu/glensdata/Individual/Q0957.html), the ratios at $x <$ 1.5 are 
not considered from here on. Using the results at $x \ge$ 1.5, we reach two conclusions: 
(a) the optical/UV ratios are not achromatic and (b) there is a correlation between $B/A$ 
and $x$, including a bump close to $x$ = 3.5, that resembles extinction laws for galaxies 
in the Local Group (Gordon et al. 2003).

\begin{figure}
\psfig{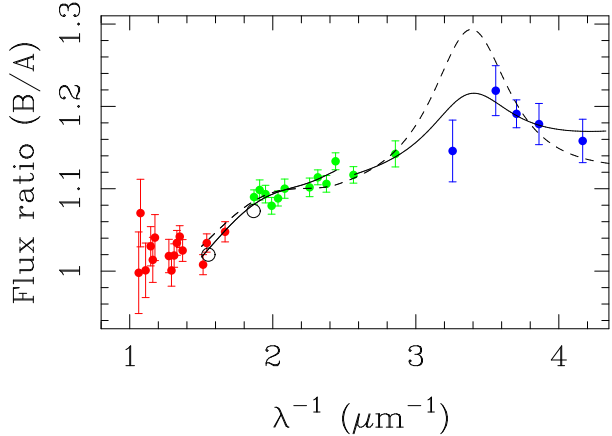}
\caption{Infrared/optical/ultraviolet continuum flux ratios of the images Q0957+561A, B. 
The blue, green and red circles are associated with the HST--STIS spectra from the G230L, 
G430L and G750L gratings, respectively. The two black open circles correspond to two 
previous measurements, and we use them to test the reliability of the new estimates at 
$x$ = 1/$\lambda$($\mu$m) $\ge$ 1.5. The dashed curve shows the best fit from Milky 
Way--like dust at the redshift of the lens galaxy, and the solid curves show the best fit 
from Local Group--like dust in the lens galaxy.}
\label{Fig. 1}
\end{figure}

Are our results consistent with only extinction and no gravitational microlensing (i.e., 
no gravitational effects due to collapsed objects in the lens galaxy)?. Radio signals 
come from regions much larger than the optical/UV continuum source, so microlensing is not 
expected to affect those signals. The extinction by dust is also irrelevant for radio 
fluxes. The flux ratio at radio wavelengths is thus assumed to be a ratio free of 
perturbations caused by microlensing and extinction, and we take $B/A$ = 0.75 (radio 
ratio) as the macrolens ratio (e.g., Garrett et al. 1994). We then fit the 20 data at $x 
\ge$ 1.5 to the Milky Way (MW)--like extinction law (Cardelli, Clayton \& Mathis 1989; 
Falco et al. 1999), varying the possible redshift of the dust system: $z_{dust}$ = 0 (MW), 
0.36 (lens galaxy), 1.125 (Lyman limit system), 1.4 (damped Ly$\alpha$ system and quasar's 
host galaxy). Details on the two Lyman systems can be found in Michalitsianos et al. (1997). 
The best fit is plotted in Fig. 1 (dashed line) and it corresponds to a dust system in the 
lens galaxy. There is no surprise: the observed bump is placed at $\lambda_{gal} \approx$  
0.21 $\mu$m, i.e., a wavelength close to the centre of the well-known extinction feature 
for local galaxies (2175 \AA). As our best solution does not accurately trace the 
observed trend ($\chi^2$ = 1.8), we use a more general extinction model. At wavelengths 
larger than 3000 \AA, the extinction curves for lines of sight in the Local Group (MW and 
Magellanic Clouds) are well described by a MW-like extinction law, whereas at wavelengths 
shorter than 3000 \AA, the extinction curves follow a model including a linear background 
term, a Drude profile and a far-UV curvature term (Gordon et al. 2003). Therefore, we fit 
our 13 measurements at $\lambda_{gal} >$ 0.3 $\mu$m to a MW model, and derive a 
differential extinction $\Delta E(B - V)$ = 67.5 $\pm$ 5.0 mmag and a ratio of total 
to selective extinction in the $V$ optical band $R_V$ = 4.4 $\pm$ 0.5 (95\% confidence 
intervals). We also fit the ratios at $\lambda_{gal} <$ 0.3 $\mu$m to a background + Drude 
model (we do not need to include the far-UV curvature), setting the Drude parameters to 
$x_0$ = 4.6 and $\gamma$ = 1. The two new fits are depicted in Fig. 1 (solid lines). They 
are characterized by $\chi^2$ values of 0.7--0.8, which suggests a slight overestimation 
of errors. We conclude that extinction by dust is a sufficient mechanism for originating 
the optical/UV continuum flux ratios and no microlensing is needed. This lack of 
microlensing signal seems to be in agreement with a small fraction of mass in collapsed 
(luminous and/or dark) structures, since the absence of microlensing signatures would be 
unlikely for a halo with a significant fraction of mass in collapsed objects (Goicoechea et
al. 2005). Previous studies in the time domain (Refsdal et al. 2000; Wambsganss et al. 
2000; Gil--Merino et al. 2001) also ruled out an important population of collapsed objects 
with substellar mass (for a radial size of the continuum emission region $R_{CER} \le$ 
10$^{-3}$ pc).

\subsection{Emission lines}

\begin{table}
 \caption{Analysis of the emission lines.}
 \begin{tabular}{@{}lcccc}
  Lines & $Contleft$ (\AA) & $Contright$ (\AA)
        & $Lineinteg$ (\AA) & $Channels$ \\
  Ly$\alpha$ & 2760--2880 & 3020--3100 & 2930--2950 & 13 \\
  \hbox{N\,{\sc v}} & 2760--2880 & 3020--3100 & 2980--3000 & 13 \\
  \hbox{C\,{\sc iv}} & 3500--3640 & 3860--3960 & 3680--3780 & 36 \\
  \hbox{C\,{\sc iii}} & 4350--4500 & 4700--4850 & 4550--4650 & 37 \\
  \hbox{Mg\,{\sc ii}} & 6500--6660 & 6820--7000 & 6720--6800 & 17 \\
 \end{tabular}

 \medskip
 $Contleft$ and $Contright$ are two continuum zones close to the lines, 
 which are used to deduce quadratic fits. From the subtraction of these 
 fits, we infer the emission line profiles in Figure 2. The lines are
 integrated over the intervals $Lineinteg$ that include 13--37 channels. 
\end{table}

\begin{figure}
\psfig{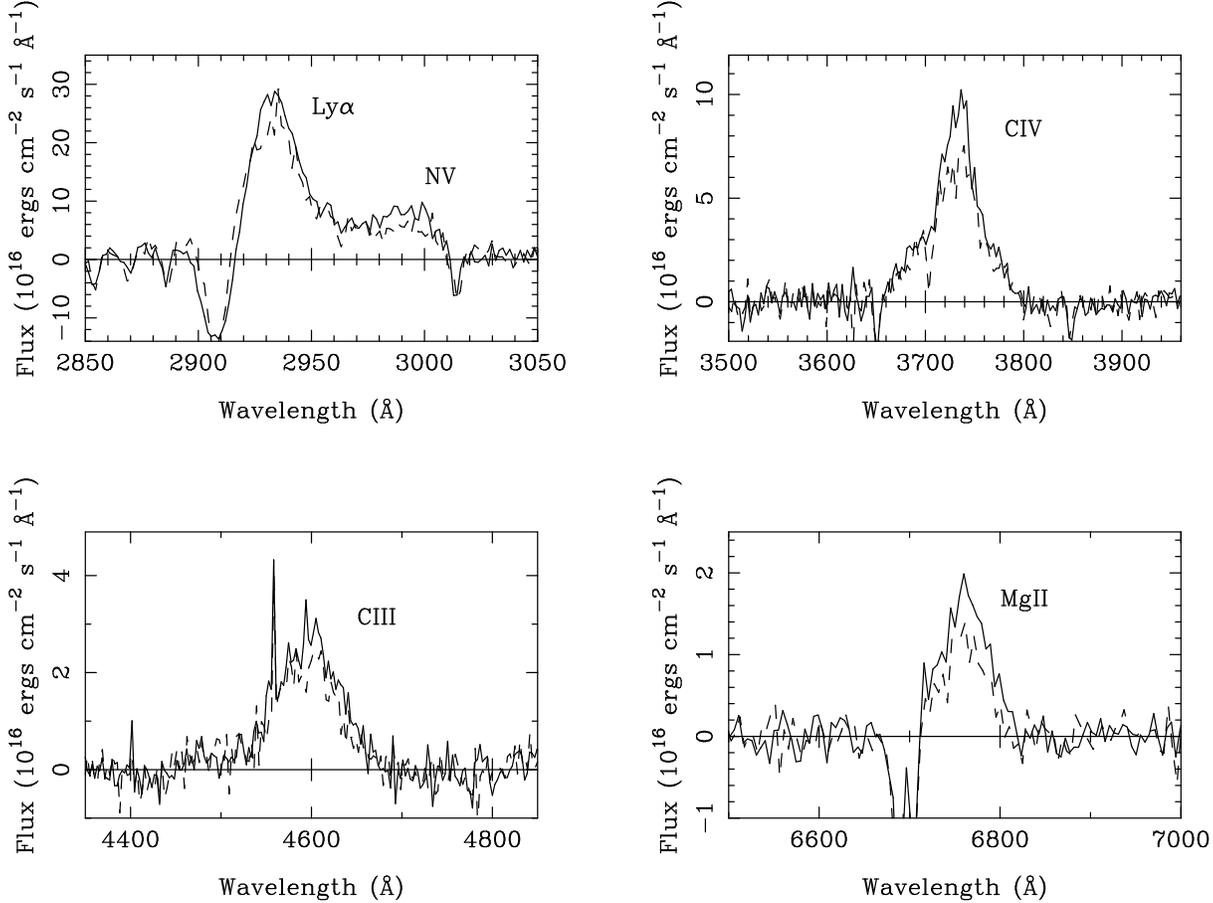}
\caption{Profiles of the main emission lines in the spectra of Q0957+561A, B. While the 
solid curves trace the features in the spectra of image A, the dashed lines describe the 
features in the spectra of image B. The profiles for the Ly$\alpha$ ($\lambda$1216) + 
\hbox{N\,{\sc v}} ($\lambda$1240) lines, the \hbox{C\,{\sc iv}} ($\lambda$1549) lines, 
the \hbox{C\,{\sc iii}} ($\lambda$1909) lines and the \hbox{Mg\,{\sc ii}} ($\lambda$2798) 
lines appear in the top left, top right, bottom left and bottom right panels, respectively. 
Emission line flux ratios ($B/A$) are less than 1, in disagreement with the continuum 
ratios in the range 2900--6800 \AA\ (1.5 $< x <$ 3.5).}
\label{Fig. 2}
\end{figure}

The HST--STIS spectra also contain several emission lines at different wavelengths. Together 
with the well-known Ly$\alpha$ and \hbox{N\,{\sc v}} (blended) ultraviolet lines (Gondhalekar
\& Wilson 1980), and \hbox{C\,{\sc iii}} and \hbox{Mg\,{\sc ii}} optical lines (Wills \& Wills 
1980), we clearly detect emission from \hbox{C\,{\sc iv}} (around 3730 \AA). As far as we know,
the \hbox{C\,{\sc iv}} lines are detected for the first time in this system. Only the reddest 
emission (\hbox{Mg\,{\sc ii}}) was properly analyzed in a previous work and a flux ratio 
$B/A$(\hbox{Mg\,{\sc ii}}) = 0.75 $\pm$ 0.02 was reported (Schild \& Smith 1991). In this paper, 
we estimate a whole set of emission line ratios, covering optical/UV wavelengths and different 
degrees of ionization. Details on the analysis of the spectral lines are given in Table 1. To
obtain each emission line profile we subtract a quadratic interpolation of the continuum, using 
two continuum zones close to the line, i.e., a zone to the left ($Contleft$ in Table 1) and 
a zone to the right ($Contright$ in Table 1). In the four panels of Figure 2 we show the profiles 
of the emission lines. After that subtraction, we integrate the resulting peak of flux over the 
interval $Lineinteg$ (see Table 1). In Table 1 we also show the number of channels corresponding 
to the integration interval ($Channels$). Due to the absorption features along the left wings of 
the Ly$\alpha$ emission lines, we only integrate the right wings of these lines. To avoid the 
possible contamination by the \hbox{N\,{\sc v}} emission, a small interval of 20 \AA\ is 
considered. The weak emission from \hbox{N\,{\sc v}} is also integrated over a 20 \AA\ interval. 
This small zone around 2990 \AA\ is not influenced by the strong  Ly$\alpha$ emission or the 
absorption at 3015 \AA. The carbon emission lines (\hbox{C\,{\sc iv}} and \hbox{C\,{\sc iii}}) 
are integrated over broader intervals of 100 \AA, whereas the right wings of the \hbox{Mg\,{\sc 
ii}} emission lines are the main contributions to their fluxes. For these lines (\hbox{Mg\,{\sc 
ii}}) we must avoid the strong absorption features around 6700 \AA. 

Our determinations of $B/A$ are: \hbox{Mg\,{\sc ii}} ($\lambda$2798) = 0.64 $\pm$ 0.04, 
\hbox{C\,{\sc iii}} ($\lambda$1909) = 0.78 $\pm$ 0.03, \hbox{C\,{\sc iv}} ($\lambda$1549) = 0.77 
$\pm$ 0.02, \hbox{N\,{\sc v}} ($\lambda$1240)  = 0.67 $\pm$ 0.07 and Ly$\alpha$ ($\lambda$1216) = 
0.87 $\pm$ 0.03. Three important conclusions arise from these results. First, we do not obtain 
any fair correlation between the $B/A$ values and the wavelength/degree of ionization. Second, 
the average of the five measurements is $\approx$ 0.75, i.e., totally consistent with the 
macrolens ratio. Third, for each pair of lines, there are several individual channels (at some 
wavelengths within the integration interval) leading to flux ratios in disagreement with the 
macrolens ratio. Therefore, it is not surprising to infer an anomalous flux ratio  
from a relatively small collection of channels. For example, due to the resolution of the 
gratings and the presence of prominent absorption features and blending, the \hbox{Mg\,{\sc 
ii}}, \hbox{N\,{\sc v}} and Ly$\alpha$ lines are studied through 13--17 channels. However, we 
use 36--37 channels for the \hbox{C\,{\sc iii}} and \hbox{C\,{\sc iv}} lines (see Table 1). We 
interpret the continuum/emission lines results in the following way: while the broad--line 
emission region (BLER) does not experience differential extinction as a whole, the continuum 
source and some substructures of the BLER do suffer it. Thus, a network of compact dusty clouds 
in the lens galaxy seems to be involved. The long--timescale evolution of $B/A$ in the $R$ 
optical filter agrees with our interpretation (Oscoz et al. 2002). The lack of microlensing in 
the continuum ratios suggests that no stars are present within the dusty regions crossing the 
A and B images. Hence, the clouds do not seem to be associated to stars and the network is 
probably embedded in the elliptical galaxy dark halo. 

Spectroscopy and multiband photometry of lensed quasars are throwing light on the 
differential extinction and microlensing of the continuum and emission line regions. Apart 
from our conclusions on the CER and BLER of QSO 0957+561 (there is differential extinction of 
the CER and some substructures of the BLER, but no gravitational microlensing) there are 
other very recent results on the subject. For example, Wucknitz et al. (2003) analyzed data of 
QSO HE 0512$-$3329. Assuming that the emission line flux ratios are only affected by differential 
extinction, the authors properly corrected the continuum flux ratios and found evidence for 
a microlensed CER. Wayth, O'Dowd \& Webster (2005) also reported flux ratios of QSO 2237+0305. 
After applying corrections for differential extinction, they argued that both the CER and 
BLER must be microlensed. We note that the four images of the system cross the bulge of a 
face--on Sab spiral galaxy. Finally, from data of QSO SBS 0909+532, Mediavilla et al. (2005) 
found differential extinction of both the CER and BLER as well as achromatic microlensing of 
the CER. 

\section{Conclusions}

We conclude that spectroscopic observations of the first gravitationally lensed quasar provide 
valuable information on the structure of the main lens (cD) galaxy at $z$ = 0.36. The data support 
the existence of a dark halo that mainly consists of non-collapsed material, rather than a 
granular dark halo harbouring a significant population of collapsed objects. The dark halo 
probably includes compact dusty clouds of gas with projected (into the source plane) radial 
sizes verifying the constraint $R_{CER} < R_{cloud} < R_{BLER}$. For example, a typical radius
$R_{cloud} \approx 10^{-2}$ pc is in agreement with measurements of $R_{CER}$ and $R_{BLER}$
of quasars (e.g., Yonehara 2001; Kaspi et al. 2000). The clouds must also be sufficiently 
diffuse that they do not appreciably gravitationally microlens the quasar light (e.g., Kerins, 
Binney \& Silk 2002). In this paper we consider a network of clouds of similar composition. 
However, a more complicated scenario incorporating different kinds of dust cannot be ruled 
out (e.g., McGough et al. 2005). Previous spectroscopic studies of the internal region of the 
cD galaxy suggested the possible existence of a central massive dark object (Mediavilla et al. 
2000), so that the new results from HST--STIS spectra of the quasar images (crossing the 
galaxy halo) complement the previous information on the galaxy nucleus. 

\section*{Acknowledgments}

We thank P. Goudfrooij for helpful comments on the CALSTIS pipeline software. Based on 
observations made with the NASA/ESA Hubble Space Telescope, obtained from the data archive at 
the Space Telescope Science Institute. STScI is operated by the Association of Universities 
for Research in Astronomy, Inc. under NASA contract NAS 5-26555. This work was supported by 
Universidad de Cantabria funds and the MCyT grant AYA2004-08243-C03-02.

\bsp

\end{document}